# Continuous-wave 148.4 nm Generation in Quasi-phase Matched $BaMgF_4$

Yuanzhuang Bu[1,2,3]†, Xiangting Wang[4]†, Ruiyin Gu[1,2,3]†, Chongtai Liu[1,2,3], Luxin Jiang[1,2,3], Xinqing Ding[5], Ke Zhang[1,2,3]*, Yongkai Li[4]*, Kemi Xu[1,2,3,4,5]*

**Affiliations:**

[1]Center for Photonic Quantum Precision Measurement, School of Interdisciplinary Science, Beijing Institute of Technology, Beijing 100081, China

[2]State Key Laboratory of Environment Characteristics and Effects for Near-space, Beijing Institute of Technology, Beijing 100081, China

[3]MIIT Key Laboratory of Complex-field Intelligent Sensing, School of Optics and Photonics, Beijing Institute of Technology, Beijing 100081, China

[4]Yangtze Delta Region Academy of Beijing Institute of Technology (Jiaxing), Jiaxing 314019, China

[5]Beijing Institute of Technology Zhuhai, Zhuhai 519088, China

*Corresponding author. Email:

kezhangvv@bit.edu.cn; yongkaili@bit.edu.cn; xukemi@bit.edu.cn

†These authors contributed equally to this work.

**Abstract**

We report continuous-wave (CW) vacuum-ultraviolet (VUV) generation at 148.4 nm by second-harmonic generation of 296.8-nm fundamental light in quasi-phase matched (QPM) $BaMgF_4$ (BMF). At the incident power of 170 mW, a VUV output power of approximately 31 pW can be obtained. The quadratic dependence of the signal on the incident power in the QPM region and its attenuation in the ambient environment support the identification of VUV generation. The polarization dependence shows a sine-function shape with a 15% output attenuation from the maximum to the minimum. These results highlight the potential of BMF as a nonlinear medium for CW VUV generation.

## Introduction

The low-energy nuclear transition of $^{229}$Th, at approximately 8.4 eV (148.4 nm), offers a unique opportunity for developing optical nuclear clocks and advancing precision nuclear spectroscopy[1-4]. High-resolution spectroscopy and coherent excitation of this transition require narrow-linewidth, frequency-stable CW VUV radiation with sufficient optical power. Developing high-performance 148.4 nm CW VUV sources is therefore essential for advancing $^{229}$Th nuclear-clock research.

Current approaches to generating 148.4-nm CW radiation primarily involve four-wave mixing in atomic vapors or second-harmonic generation (SHG) in solid-state nonlinear crystals. In 2026, Xiao et al. demonstrated resonantly enhanced four-wave mixing in cadmium vapor, achieving over 100 nW of CW output at 148.4 nm[5]. Subsequently, Huang et al. increased the output power to approximately 10 μW and demonstrated the operation of a $^{229}$Th nuclear clock[3]. However, vapor-based frequency conversion requires precise control of atomic resonance and vapor conditions. Solid-state nonlinear frequency conversion offers an alternative route to VUV generation. In 2025, Lal et al. demonstrated random QPM SHG in strontium tetraborate ($SrB_4O_7$, SBO), generating approximately 1.3 nW at 148.4 nm with 325 mW of incident fundamental power[6]. In 2026, Subramanian et al. achieved approximately 16 pW at 148.4 nm through cavity-enhanced SHG in periodically poled BMF[7]. Unlike random quasi-phase matching, periodic poling enables controlled compensation of phase mismatch through spatial modulation of the second-order nonlinear coefficient. Nevertheless, the reported BMF-based VUV output remains at the picowatt level, motivating further improvements in frequency-conversion performance.

Here, we investigate SHG from 296.8 to 148.4 nm using a QPM BMF crystal. By optimizing the QPM SHG process, we achieve approximately 31 pW of CW VUV output. This work demonstrates the potential of BMF for enhanced VUV frequency conversion and advances the development of solid-state CW sources for precision spectroscopy of the $^{229}$Th nuclear transition and future optical nuclear clocks.

## BMF Crystal Preparation and Characterization

The BMF single crystal used in this study was grown by the Czochralski pulling method from high-purity stoichiometric mixtures of $BaF_2$ and $MgF_2$ precursors[8]. The growth process was carried out under a protective inert atmosphere, with precise control over the pulling rate, crystal rotation speed, and axial temperature gradient to suppress spontaneous nucleation and reduce crystalline defects, yielding centimeter-scale single crystals with high optical transparency. Fig. 1(a) schematically illustrates the orthorhombic crystal structure of BMF, in which $Ba^{2+}$ cations occupy high-coordination lattice sites and $Mg^{2+}$ cations are tetrahedrally coordinated by $F^-$ anions to form discrete $MgF_4$ units. These structural units are interconnected via bridging fluoride ions to build an anisotropic three-dimensional framework, giving rise to the birefringence required for phase-matched vacuum-ultraviolet frequency conversion[9]. The as-grown crystal is colorless and optically transparent; a photograph of a fabricated sample is displayed in the inset of Fig. 1(b), showing well-defined geometry and good macroscopic optical quality.

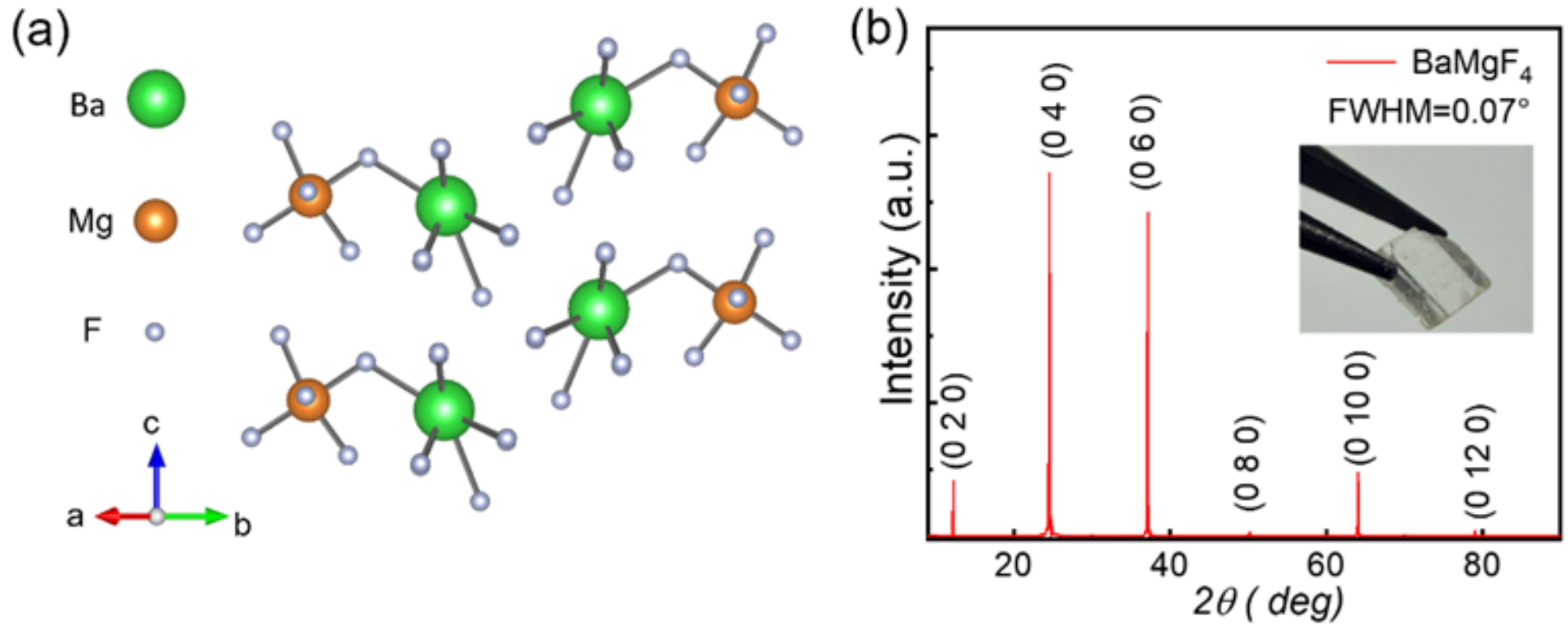

**Fig. 1** (a) Schematic diagram of crystal structure of BMF. (b) XRD pattern of BMF single crystals with (0 1 0) reflections, and the inset displays a photograph of the cut crystal.

A (010)-oriented specimen was cut and double-side polished from the bulk crystal for structural characterization and subsequent nonlinear optical experiments. Fig. 1(b) presents the X-ray diffraction (XRD) pattern of the oriented BMF sample, which features only a series of sharp (0 k 0) reflections (k = 2, 4, 6, 8, 12) with no detectable peaks from impurity phases or misoriented crystallites. The exclusive presence of (0 k 0) diffractions confirms the single-crystalline nature and excellent out-of-plane orientation of the prepared specimen. The (0 4 0) reflection exhibits a full width at half maximum (FWHM) of merely 0.07°, indicative of high crystalline perfection and low residual lattice strain. In contrast to domain-structured nonlinear crystals that typically show peak splitting and angular shifts arising from local lattice variations at domain boundaries, the symmetric and narrow diffraction peaks observed here imply uniform interplanar spacing and homogeneous structural properties throughout the crystal volume. According to Bragg's law, the measured diffraction angles are in good agreement with the standard lattice parameters of orthorhombic BMF, verifying the phase purity and structural uniformity of the Czochralski-grown crystal, which meets the requirements for high-performance vacuum-ultraviolet nonlinear optical materials.

The BMF crystal used for the VUV generation has the size of $3.2 \times 1.2 \times 2.7$ $mm^3$. Transmission spectrum of BMF crystal along the b axis is measured with a deuterium lamp and a VUV spectrometer without a polarizer under vacuum (as shown in Fig. 2).

The sample with high crystallinity shows a relatively low VUV transmission (e.g., lower than 10% at the wavelength ranging from 143 nm to 153 nm, as shown in the inset of Fig. 2), which is ascribed to the inevitable scattering centers[7]. The transmission of the 1.2 mm thick sample is 6.8% at the wavelength of 148.4 nm, giving an absorption coefficient of approximately 22 $cm^{-1}$.

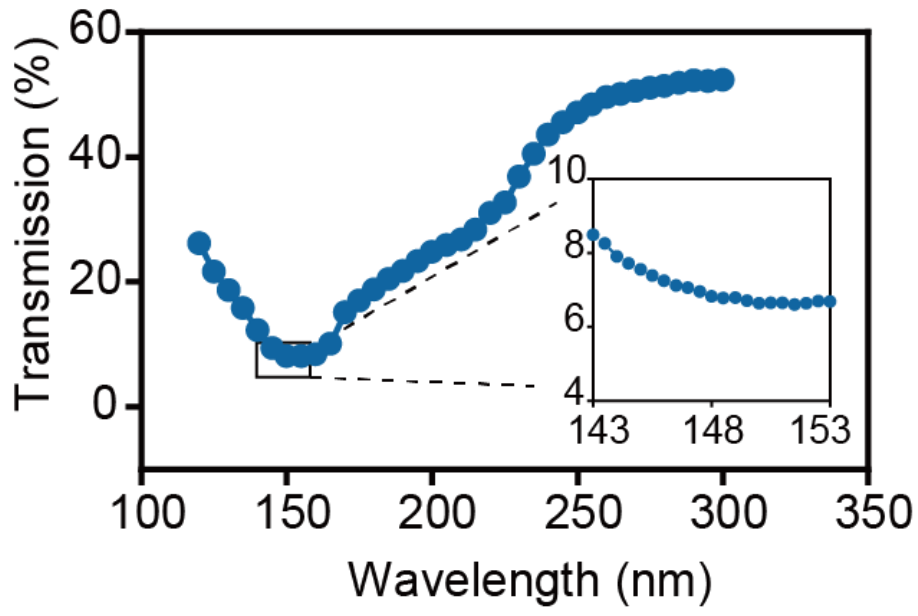

**Fig. 2** Transmission spectrum of the BMF crystal.

## SHG Generation and Detection Setup

Fig. 3 shows the experimental setup for CW SHG at 148.4 nm based on a BMF crystal. The fundamental light is provided by a commercial CW laser (Precilasers, FL-SF-297-1-CW) operating at 296.8 nm, with a maximum output power of 1130 mW and a specified linewidth of ~15.8 kHz. The fundamental light is incident on the BMF crystal along the direction of b axis. A two-lens optical system is employed to optimize the beam waist and focusing conditions, while a half-wave plate and a Glan-Laser prism are used to control the polarization state of the fundamental light. The nonlinear frequency conversion process is carried out in a vacuum chamber. The BMF crystal is mounted on a motorized translation stage, allowing precise adjustment of its position to optimize the nonlinear interaction. A $CaF_2$ prism is used to separate the generated VUV signal from the fundamental beam and direct the VUV radiation to a PMT (Hamamatsu R1080) for power detection.

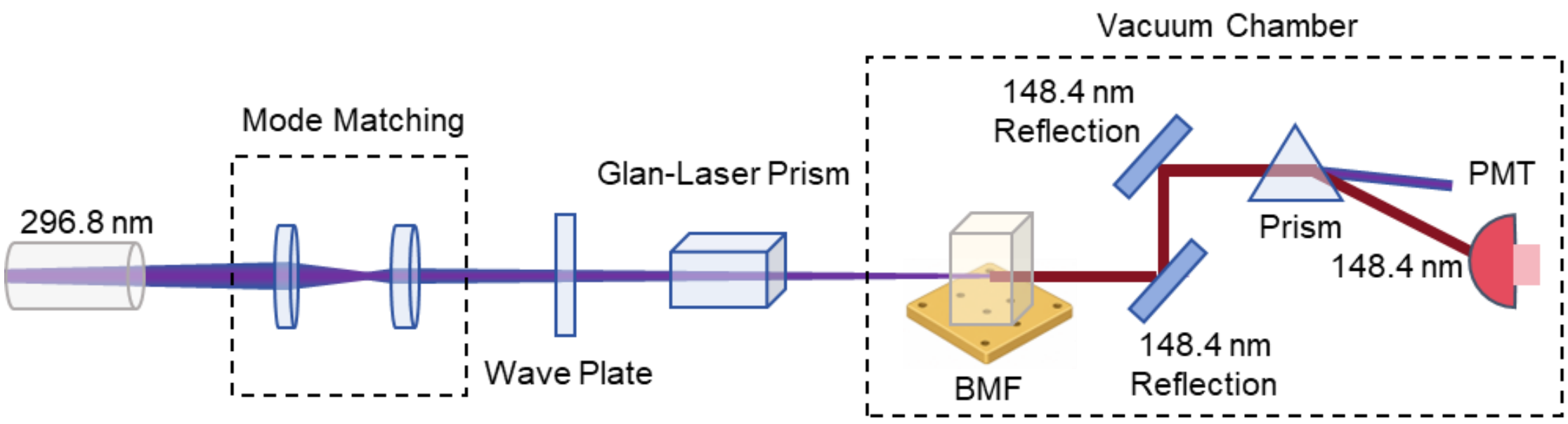

**Fig. 3** Experimental setup for CW SHG at 148.4 nm. The 296.8 nm CW laser is subjected to beam optimization and polarization adjustment before being directed into the BMF crystal to generate

VUV light. A $CaF_2$ prism separates the VUV signal from the fundamental light, and a PMT is used to measure the VUV power.

## Identification and Characterization of VUV SHG

To investigate VUV generation, we measured the PMT signal, $P_{PMT}$, as a function of the fundamental power, $P_F$. As shown in Fig. 4, the data exhibit a quadratic dependence on $P_F$ at the phase-matching region in the vacuum environment. The inset of Fig. 4 demonstrates that the PMT signal decreases monotonically with the increasing chamber pressure as air is gradually introduced into the chamber. Taken together, these experimental observations support the attribution of quadratic signal component to VUV SHG, which is absorbed intensely in the ambient environment. The residual signal at atmospheric pressure is attributed to background from reflected and scattered fundamental light. By moving the incident beam slightly away from the QPM region to the region without QPM, $P_{PMT}$ decreases rapidly and exhibits an approximately linear dependence on $P_F$, which is consistent with the scattering of fundamental pump light. The VUV power at the BMF crystal exit, $P_{BMF}$, can be estimated as follows[10]:

$$P_{PMT} = aP_F^2 + bP_F + c \tag{1}$$

$$P_{BMF} = aP_F^2/(\eta_{148.4} \times R_1 \times R_2 \times T_{prism} \times \varepsilon_{PC} \times \varepsilon_{Opt}) \tag{2}$$

In equation (1), a,b and c are the fitting coefficients, respectively. In equation (2), $\eta_{148.4}$ is the quantum efficiency of PMT at 148.4 nm, $R_1$ and $R_2$ are the reflectance of the two VUV mirrors, $T_{prism}$ is the efficiency of the triangular $CaF_2$ prism, $\varepsilon_{PC}$ is the photoelectron collection and pulse-discrimination acceptance, and $\varepsilon_{Opt}$ is the optical collection efficiency.

To quantitatively estimate the $P_{BMF}$, here we take $\eta_{148.4}$=0.08, $R_1 \approx R_2 \approx 0.95$, $T_{prism} \approx 0.4$ and assume that $\varepsilon_{PC} = \varepsilon_{Opt} = 1$. From the data in Fig. 4, it can be derived that $a = 3.1\times10^{-14}$ mW$^{-1}$. At an incident power of 170 mW, it can be estimated that $P_{BMF} \approx 31$ pW.

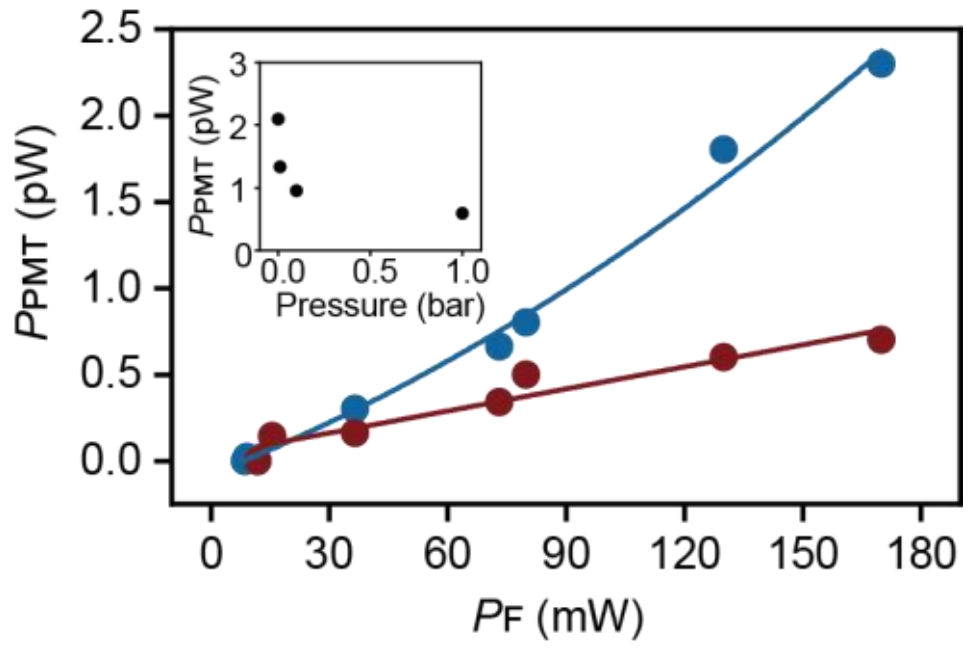

**Fig. 4** PMT signal as a function of the 296.8 nm fundamental incident power on the BMF crystal. Blue and red points mark the measurement in the QPM region and the region without QPM. Blue line shows a quadratic-polynomial fit to the QPM region data. Red line shows a linear fit to the region without QPM. The inset shows the dependence of PMT signal on the environment pressure.

We further investigate the polarization dependence of VUV SHG signal. A $\lambda/2$ waveplate and a Glan-Laser prism, both positioned upstream of the vacuum chamber, are used to control the polarization of the fundamental pump beam incident on the BMF crystal. Fig. 5 shows the measured signal as a function of the incident polarization angle. The signal reaches its maximum at 90°. The signal decreases by ~15% compared with the maximum and reaches its minimum at 0° and 180°.

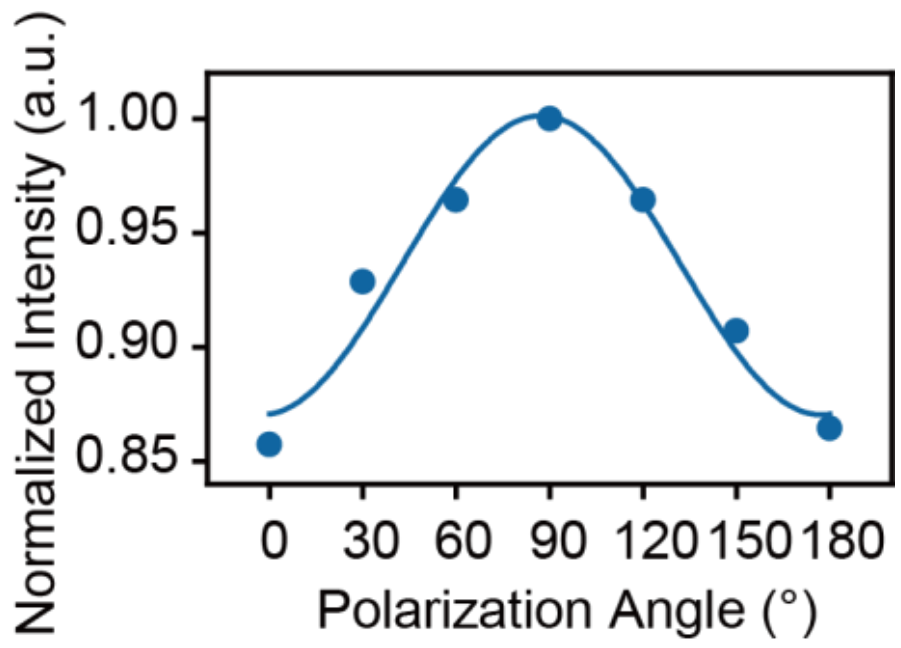

**Fig. 5** Polarization dependence of the VUV SHG signal.

## Conclusion

In this article, we report CW VUV generation at the wavelength of 148.4 nm by the SHG in QPM BMF crystal. The VUV generation is detected in a vacuum chamber with a PMT, showing quadratic dependence on incident pump power and intense suppression in ambient environment. An approximately 31 pW output power is obtained at the crystal exit with an incident power of 170 mW. The signal reaches its maximum at the polarization angle of 90° and decreases by approximately 15% for the orthogonal polarization. These results provide an experimental basis for further development of CW VUV sources based on QPM fluoride crystals.

## Funding and Acknowledgements

This work is supported by Fundamental and Interdisciplinary Disciplines Breakthrough Plan of the Ministry of Education of China. We thank Linqiang Hua of Innovation Academy for Precision Measurement Science and Technology, Chinese Academy of Sciences, for providing optical components for VUV detection. We also thank Zhiqiang Zhang of University of Science and Technology of China for helpful suggestions on the manuscript.